\documentclass[preprint,12pt,authoryear]{elsarticle}

\usepackage{amssymb}
\usepackage{amsmath}
\usepackage{booktabs}
\usepackage{float}
\usepackage{array}
\usepackage{tikz}
\usepackage{graphicx}
\usepackage{tcolorbox}
\usepackage{longtable}
\usepackage{xurl}

\begin{document}

\begin{frontmatter}

\title{The AI Literacy Heptagon: A Structured Approach to AI Literacy in Higher Education}

\author[1]{Veronika Hackl\corref{cor1}}
\author[2]{Alexandra Elena Müller}
\author[1]{Maximilian Sailer}

\address[1]{Faculty of Social and Educational Sciences, University of Passau, Germany}
\address[2]{Faculty of Law, University of Passau, Germany}

\cortext[cor1]{Corresponding author: veronika.hackl@uni-passau.de}

\end{frontmatter}

\begin{abstract}
The integrative literature review addresses the conceptualization and implementation of AI Literacy (AIL) in Higher Education (HE) by examining recent research literature. Through an analysis of publications (2021-2024), we explore (1) how AIL is defined and conceptualized in current research, particularly in HE, and how it can be delineated from related concepts such as Data Literacy, Media Literacy, and Computational Literacy; (2) how various definitions can be synthesized into a comprehensive working definition, and (3) how scientific insights can be effectively translated into educational practice. Our analysis identifies seven central dimensions of AIL: technical, applicational, critical thinking, ethical, social, integrational, and legal. These are synthesized in the AI Literacy Heptagon, deepening conceptual understanding and supporting the structured development of AIL in HE. The study aims to bridge the gap between theoretical AIL conceptualizations and the practical implementation in academic curricula.
\end{abstract}


\section{Introduction}
\label{sec1}

\begin{quotation}
``You don't have to burn books to destroy a culture. Just get people to stop reading them.''
\begin{flushright}
--- Ray \citet{Bradbury1953}, \textit{Fahrenheit 451}
\end{flushright}
\end{quotation}

The rapid integration of Artificial Intelligence (AI) into professional and civic life requires an expansion of the concept of literacy itself. Societal participation now increasingly depends on the ability to critically engage with complex AI systems. This emerging skill, termed AI Literacy (AIL), is broadly defined as the competencies to “critically evaluate AI technologies; communicate and collaborate effectively with AI; and use AI as a tool online, at home, and in the workplace” \citep{Long2020}.

Fostering these competencies is now considered a crucial mission for Higher Education (HE) in preparing students for an AI-driven future \citep{Chai2024, KitNg2022, Miao24, Schmohl2023}. The consensus on its importance is underscored by its designation as a key 21st-century competency \citep{Almatrafi2024, Long2020, Schmohl2023, Wannemacher2021} and by policy mandates, such as the EU AI Act's explicit call for AIL promotion \citep{EU}.

Despite this consensus, the systematic integration of AIL into HE curricula remains notably limited. Evidence from Germany, for instance, shows that only 31.6\% of students report that AIL is explicitly addressed in their programs \citep{BuddeFriedrich2024, Lambrecht2025}. A primary obstacle to closing this implementation gap is the conceptual ambiguity of AIL itself. While universities are tasked with this mission \citep{Knoth2024, Miao24, Wannemacher2021, Delcker2024}, vague definitions and a lack of operational frameworks lead to a fragmented transfer into practice.

To address these challenges, this study follows a structured, multi-stage approach grounded in an integrative literature review. We begin by establishing a theoretical foundation, analyzing how AI Literacy is currently conceptualized in research literature and delineating it from related literacies such as Data and Media Literacy (RQ1). Building upon this analysis, we synthesize the findings into a comprehensive working definition that provides a clear conceptual anchor for our work (RQ2). The primary contribution of this paper is then to translate these theoretical insights into a practical instrument for educational practice: the 'AI Literacy Heptagon'. This structured, seven-dimensional framework is designed to support systematic curriculum development and analysis (RQ3).
To test the framework's utility, we conduct an initial validation by mapping its dimensions onto the curricula of two different academic programs. We conclude by discussing the implications of our framework for HE and suggesting critical directions for future research.

\section{Current State of Research and Research Questions}
\label{sec2}

Before proposing a new framework, it is essential to first establish an understanding of the existing AIL landscape. This section therefore reviews the current conceptualizations of AIL, identifies research gaps, and formulates the research questions that guide our study. 

\subsection{AI Literacy conceptualizations and frameworks}

Conceptualizations of AI Literacy (AIL) have matured significantly from an early focus on purely technical understanding \citep{Burgsteiner, Kandlhofer2016} to a more holistic view encompassing complex competencies, mirroring the evolution of related fields like digital or media literacy \citep{Sperling2024}. This maturation has led to a distinction in the current literature: the separation between generic AIL for non-experts and domain-specific AIL for specialists \citep{Almatrafi2024, Long2020}. Generic AIL equips individuals with the high-level understanding needed to critically evaluate AI's societal impact and to use common applications responsibly, such as recognizing algorithmic bias or navigating data privacy \citep{Laupichler2023, Ng2021}. In contrast, domain-specific AIL integrates the technical and contextual knowledge essential for professional practice in fields like medicine or engineering, including topics like machine learning principles and regulatory considerations \citep{Knoth2024, Schleiss2023, cerny13030129}.

While the foundational definition by \citet{Long2020} provides a widely accepted starting point, several major frameworks have attempted to operationalize AIL, each with a distinct focus. Specifically for the HE context, the EDUCAUSE working group emphasizes the practical application in teaching and learning, focusing on the critical evaluation of AI tools and vigilance against ethical risks like bias and misuse \citep{educause2024ailiteracy}. From a global policy perspective, the UNESCO AI Competency Framework offers a broader view, structuring AIL around dimensions of understanding, evaluation, responsible use, and creation, while placing a strong emphasis on inclusive and equitable development \citep{Miao24}.The joint "AI Literacy Framework for Primary and Secondary Education" by the European Commission and the OECD \citep{OECD2025ailit} marks a recent reference point. Although targeted at the K-12 level, the AILit framework structures AIL around a foundation of knowledge, skills, and attitudes and defines four primary domains of interaction: Engaging with AI, Creating with AI, Managing AI, and Designing AI. The emergence of this comprehensive K-12 framework accentuates the need for a similarly structured but distinctively tailored model for the HE sector.

However, significant limitations emerge when attempting to integrate AIL into broader digital competency models. The European DigComp 2.2 framework, for example, offers numerous illustrations for citizen-AI interaction but explicitly states it is not a comprehensive curriculum and does not formalize AIL as a discrete competency, leaving key educational aspects unaddressed \citep{Vuorikari2022}. Similarly, the IEEE DQ Framework subsumes AIL within 'Digital Literacy' for decision-making but does not elaborate on the specific pedagogical dimensions required to foster these skills effectively within a university setting \citep{IEEE3527.1-2020}. The most recent Ultimately, this review reveals a landscape of valuable but fragmented perspectives, where no single framework provides a coherent, actionable synthesis for the specific needs of HE.

\subsection{Research gaps}
This synthesis of the literature brings three research gaps into focus for the HE context:
\begin{itemize}
    \item Lack of a synthesized, operational definition: Despite numerous conceptual approaches, a unified definition that integrates the multifaceted nature of AIL is lacking, specifically in HE contexts \citep{Pinski2024, mustafa2024systematic}, with \citet{Sperling2024} emphasizing the need for an education-specific definition that differentiates AIL from related literacies, such as Digital, Media, and Information Literacy.
    
    \item Absence of an actionable framework for implementation: The field lacks a structured yet adaptable framework to translate AIL concepts into educational practice. This deficit is the primary cause for the observed fragmentation in implementation: educators lack concrete guidelines \citep{Pinski2024}, and initial approaches demonstrate inconsistent scope and methodology \citep{odeang2024prelims, fleischmann2024fostering}. Without a common structure, implementations remain isolated to specific applications or disciplines \citep{Schleiss2023} instead of enabling systematic, institution-wide integration \citep{leriasinfo15040205, batistainfo15110676}.
    
    \item Neglect of integrative and regulatory dimensions: Existing conceptualizations do not address legal and integrative dimensions, particularly problematic given the rapidly evolving regulatory landscape.
\end{itemize}

\subsection{Research Questions}
Based on the identified research gaps, this study is guided by the following research questions:
\begin{itemize}
    \item RQ1: How is AIL conceptualized in the current research literature, and how can it be delineated from related literacies within the HE context?
    \item RQ2: How can a working definition capture and synthesize both recurring and emerging themes in AIL conceptualizations?
    \item RQ3: How can these conceptual dimensions be structured into a practical framework to guide systematic curriculum analysis and development in HE?
\end{itemize}

These questions are designed to progress from a comprehensive conceptual understanding (RQ1, RQ2) to the development of an actionable instrument that translates theory into practice (RQ3).

\section{Methodology}

Having established the guiding research questions, this section details the systematic approach taken to answer them. We outline the integrative literature review method used to analyze the current state of AIL and describe the expert-led curriculum mapping process employed for the initial and basic validation of our framework.

\label{sec3}
\subsection{Literature review}
\begin{figure}[H]
    \centering
    \includegraphics[width=0.8\textwidth]{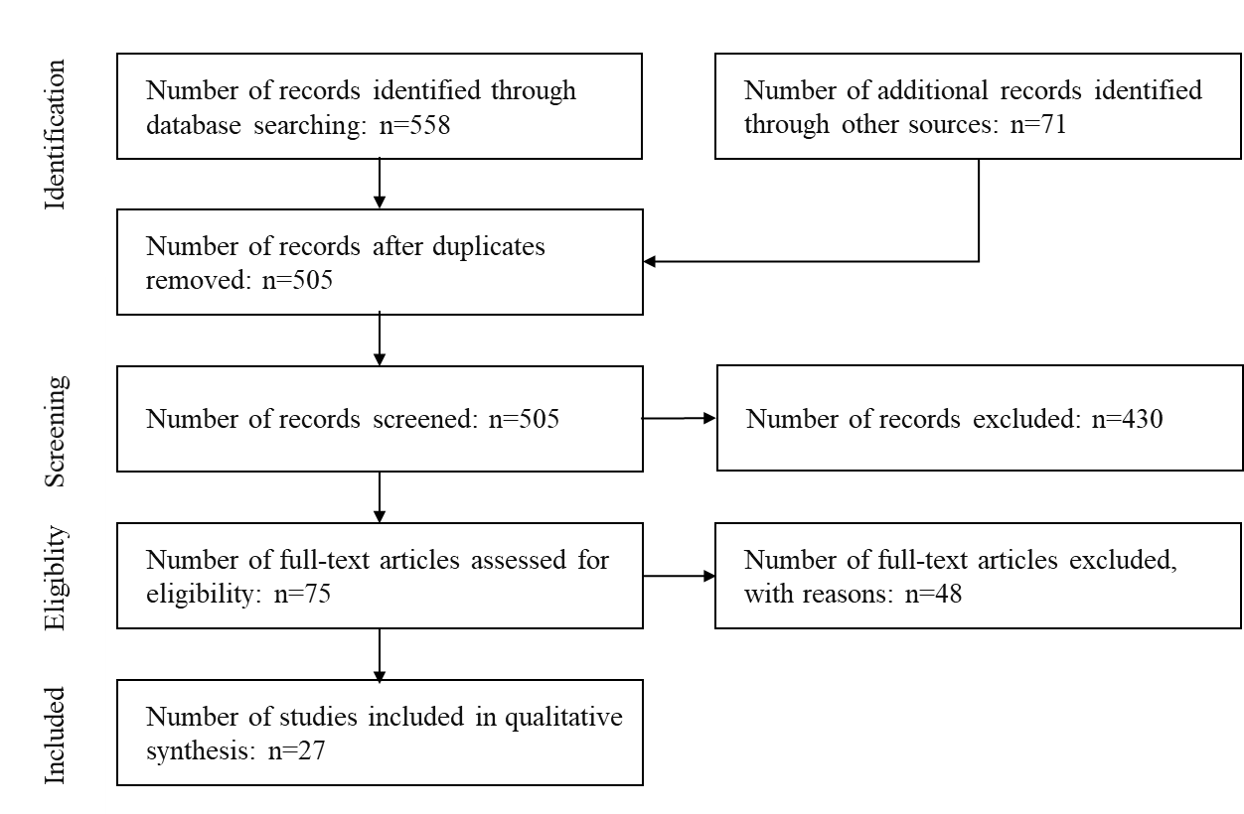}
    \caption{Review process following the PRISMA statement}
    \label{fig:meinbild}
\end{figure}

To address the research questions, the integrative literature review method \citep{Torraco2005, Snyder2019, Moher2009} was chosen as it allows for a comprehensive examination of the various aspects of AIL in HE while drawing practice-oriented conclusions, particularly with respect to the identification and validation of key themes in emerging fields where established frameworks are still evolving. This approach differs from a traditional systematic review by adopting a broader research scope and incorporating various conceptualizations. It is distinct from a meta-analysis as it does not perform a statistical synthesis of quantitative studies. Instead, the focus is on the qualitative analysis and synthesis of the literature to develop a comprehensive understanding of AIL and its practical implementation in HE. To ensure transparency and reproducibility, the study follows key principles of the PRISMA framework \citep{Moher2009}, particularly in documenting the literature selection process (see figure \ref{fig:meinbild}).
Following the PRISMA guidelines, the search process was systematically recorded, including the number of initial results, the screening of duplicates, and the final selection process. The systematic search of peer-reviewed literature was carried out using electronic scientific databases (Web of Science and Scopus). The search was limited to English-language publications from 2021 (and earlier foundational works) to 2024, focusing on final publications including articles, book chapters, and books (compare tab. \ref{tab:selection_criteria}). Search strings were developed using combinations of key terms related to AIL ("AIL," "artificial intelligence literacy," "digital literacy in AI," "AI skills," "artificial intelligence skills") and HE context ("higher education," "tertiary education," "universities," "colleges"). Additional search terms included "teaching AI," "learning AI," "AI education," "AI competencies," and "AI curriculum development" combined with stakeholder terms ("students," "faculty," "educators," "academic staff"). Additional sources were identified through citation tracking of seminal papers, e.g. \citet{Long2020}, Google Scholar searches using the string "AI Literacy in Higher Education", research discovery tools (ResearchRabbit) and Science Direct recommendations. The literature search was completed on December 10, 2024.

Table \ref{tab:thematic_coding} organizes the concept of AIL into seven distinct dimensions which emerged through an iterative coding process conducted by two independent coders. Initially, open coding was applied to extract recurring competencies and thematic elements from various definitions and conceptualizations of AIL in HE. Subsequent axial coding allowed us to group similar codes into broader themes. Discrepancies were resolved through discussion to reach a consensus. This led to the identification of traditional dimensions such as technical knowledge and skills, as well as additional dimensions, e.g., integration skills and legal and regulatory knowledge, to fully capture the multifaceted nature of AIL.

\subsection{Initial Framework Validation through Expert-Led Curriculum Mapping}

To assess the practical applicability and relevance of the proposed AI Literacy Heptagon, we conducted an initial, qualitative validation using an expert-led curriculum mapping methodology. This approach was chosen not to empirically measure student competencies, but to evaluate whether the framework could serve as a meaningful tool for analyzing and structuring existing HE curricula.

Two experts were selected to represent distinct disciplinary contexts. The first expert is the program director for a highly technical Bachelor's program in 'AI Engineering', possessing deep expertise in computer science and AI curriculum development. The second expert is the curriculum coordinator for a 'Media Pedagogy' program within a teacher education department, with extensive experience in integrating digital media and critical thinking into humanities and social science curricula. This selection allows for testing the framework's versatility across both STEM and socio-pedagogical domains.

The validation was conducted through structured "collaborative profiling sessions" with each expert individually. The process followed a defined protocol:

\begin{enumerate}
\item Framework Introduction: The researchers presented the AI Literacy Heptagon, explaining each of the seven dimensions and the four proficiency levels. The proficiency levels (Unaware, Beginner, Intermediate, Expert) were explicitly linked to the cognitive processes of Bloom's Taxonomy (see Table 3) to provide a shared pedagogical basis for evaluation.
\item Curriculum Mapping Task: The experts were asked to systematically analyze their respective curriculum documents (i.e., module handbooks, course descriptions, and defined learning outcomes) and map the existing content to the seven dimensions of the heptagon.
\item Proficiency Level Assignment: For each dimension, the experts determined the target proficiency level that a typical graduate of their program is expected to achieve. This was a qualitative judgment based on the depth and breadth of coverage in the curriculum.
\item Framework Evaluation: During the process, the experts provided qualitative feedback on the framework itself, commenting on the clarity, comprehensiveness, and relevance of the dimensions for their specific field.
\end{enumerate}

The output of this process was a visual heptagon profile for each program, representing the intended curricular focus and identifying potential competency gaps. It is crucial to note that these profiles, as shown in Figure 4, are illustrative representations of curricular goals and structure, not empirical measurements of student performance. This expert-centered mapping serves as a foundational step in validating the framework's utility for educational practice.

\section{Results}
\label{sec4}
The application of this methodology yielded the following findings. This section presents these results sequentially, structured according to our three research questions. We begin by analyzing the conceptualization of AIL in recent literature.
\subsection{Core dimensions and conceptualizations of AI Literacy (RQ1)}
\subsubsection{Core Dimensions}

The analysis reveals that recent AIL conceptualizations between 2021-2024 demonstrate a shift toward a more comprehensive approach that incorporates ethical, social, and practical dimensions, e.g., the constructs identified by \cite{Almatrafi2024}: Recognize, Know and Understand, Use and Apply, Evaluate, Create, and Navigate Ethically. We found a lack of domain-specificity, digital integration aspects, and legal and regulatory awareness in existing definitions. As Table \ref{tab:thematic_coding} indicates, Technical Knowledge and Skills (TKS), Application Proficiency (AP), and Critical Thinking Ability (CTA) emerged consistently in most of the frameworks analyzed, forming the core competency areas. Similarly, Ethical Awareness and Reasoning (EAR) was identified in a large majority of sources (20/27), highlighting its central role in responsible AI engagement.
However, we deliberately included dimensions that appeared less frequently but represent critical emerging areas. Thus, Social Impact Understanding (SIU) was incorporated despite moderate representations (11/27). Integration Skills (IS) with 18/27 sources furthermore revealed a gap between theoretical AI knowledge and practical application, a dimension increasingly important as AI systems become embedded in diverse workflows. Legal and Regulatory Knowledge (LRK) was included despite appearing in only 2/27 sources, addressing the rapidly evolving regulatory landscape exemplified by the EU AI Act (Regulation (EU) 2024/1689). This approach ensures that the framework addresses not only current conceptualizations, but also anticipates future competency requirements. Dimensions considered but ultimately integrated within our seven core areas included technical sub-skills like prompt engineering (incorporated into AP) and domain-specific applications (addressed through our generic-to-specialized progression model rather than as separate dimensions).

\begin{table}[H]
    \centering
    \renewcommand{\arraystretch}{1.2}
    \setlength{\tabcolsep}{8pt}
    \small
    \begin{tabular}{p{7cm} p{1cm} p{1.5cm}}
        \toprule
        \textbf{Dimension} & \textbf{Code} & \textbf{Score (n=27)} \\
        \midrule
        \textbf{Technical Knowledge and Skills} & TKS & 24/27 \\
        \midrule
        \textbf{Application Proficiency} & AP & 26/27 \\
        \midrule
        \textbf{Critical Thinking Ability} & CTA & 23/27 \\
        \midrule
        \textbf{Ethical Awareness and Reasoning} & EAR & 20/27 \\
        \midrule
        \textbf{Social Impact Understanding} & SIU & 11/27 \\
        \midrule
        \textbf{Integration Skills} & IS & 18/27 \\
        \midrule
        \textbf{Legal and Regulatory Knowledge} & LRK & 2/27 \\
        \bottomrule
    \end{tabular}
    \caption{Thematic Coding Framework for AIL}
    \label{tab:thematic_coding}
\end{table}

The seven dimensions explained: 

\begin{itemize}
    \item Technical Knowledge and Skills (TKS): The understanding of fundamental AI principles, including algorithms, data processing, and computational mechanisms. TKS encompasses knowledge of how AI systems process input data, learn from datasets, and generate outputs \citep{Almatrafi2024}. It includes familiarity with key AI models, their functions, and interconnections \citep{Miao24}. While programming skills are not required \citep{Mansoor2024}, a basic grasp of computational concepts such as abstraction and algorithmic thinking is essential for evaluating AI’s capabilities, limitations, and ethical implications \citep{KitNg2022}.
    \item Application Proficiency (AP): Effectively utilizing AI technologies across diverse contexts to solve practical problems and improve workflows as well as understanding various AI tools' strengths and limitations \citep{Cardon2023}. It involves the capacity to assess AI-generated outputs, refine interaction strategies, and adapt AI applications to domain-specific requirements, ensuring efficient and contextually appropriate AI deployment \citep{folmeg, Cardon2023}. AP includes, e.g., developing effective prompts for language models through prompting techniques \citep{Knoth2024, cerny13030129}, using AI writing assistants for literature review structuring or AI coding tools for data analysis in research projects.
    \item Critical Thinking Ability (CTA): The capacity to analyze, evaluate, and critically assess AI systems, their capabilities, limitations, and implications for specific use cases \citep{Tzirides2024, folmeg, Carolus2023, Long2020}. Drawing from established critical thinking frameworks \citep{lai2011critical}, it involves analyzing arguments, making inferences through reasoning, evaluating claims, and solving problems when engaging with AI technologies. This requires both background knowledge about AI and intellectual dispositions including open-mindedness, inquisitiveness, flexibility, and a willingness to seek evidence rather than accepting AI outputs uncritically. 
    \item Integration Skills (IS): The ability to effectively incorporate and adapt AI technologies into diverse digital environments and workflows is critical for navigating professional contexts increasingly transformed by (generative) AI \citep{farrellyeducsci13111109, annapureddy10.1145/3685680, Mansoor2024}. It includes the capability to identify appropriate integration points for AI tools within existing systems, modify workflows to optimize AI collaboration, and adapt to rapid technological advancements. Integration skills extend beyond mere technical implementation to include understanding organizational impacts, workflow redesign, and managing the transition between human and AI-driven processes \citep{Almatrafi2024}.
    \item Legal and Regulatory Knowledge (LRK):
   The regulation of AI is a highly dynamic and globally fragmented field. The legal landscape is shaped by two distinct, but interacting forces. On one hand, it is the creation of new, AI-specific legislation and on the other hand, the application of already existing technology-neutral legal frameworks. A comprehensive AI Literacy must account for both, as their interplay defines the current regulatory reality, which is characterized by significant "cross-jurisdictional inconsistencies" \citep{Hacker2025OxfordHandbook}.
The most prominent example of current dedicated AI regulations is the European Union's AI Act (Reg 2024/1689) \citet{EU}, and "European users should become familiar with it and its implications", as \citet{annapureddy10.1145/3685680} puts it. To support this, sector-specific education and training should be strengthened, as \citet[p. 37]{hacker2024} recommends, although this relates to primarily regulated industries in the working environment, and not HE. The AI Act is considered the first comprehensive measure of its kind with wide applicability and a risk-based approach. This contrasts with approaches in other regions. In the United States, the regulatory environment is described as an "intricate patchwork of federal initiatives and state level regulations," leading to a dynamic landscape where individual states address specific harms, such as Tennessee's laws protecting against unauthorized AI-generated likenesses or Colorado's transparency mandates. This trend of crafting specific national or regional rules is a global phenomenon, with countries like South Korea and Brazil also enacting their own distinct legislative frameworks \citep{Hacker2025OxfordHandbook}.
Furthermore, the existing legal regimes that were not designed for AI but are directly applicable and often with more immediate impact. Foundational domains such as data protection (e.g., the GDPR), copyright, and competition law are being actively enforced in the context of generative AI \citep{Hacker2025OxfordHandbook}. High-profile lawsuits over the use of copyrighted training data and antitrust investigations into major AI partnerships underscore that AI does not operate in a legal vacuum. As \citet{Hacker2025OxfordHandbook} conclude in their analysis, these established frameworks currently regulate generative AI "more tightly and effectively" than newer, AI-specific laws that are still being fully implemented.
Therefore, the LRK dimension of AI Literacy requires a multifaceted understanding. One must be aware of their region's AI-specific act. \citet[p.80]{Vuorikari2022} states that individuals should understand the privacy, security, and legal risks of AI systems that process personal or biometric data, and be able to weigh the benefits and risks of their use. A literate individual must appreciate the global regulatory diversity and the dynamic interplay between these regulatory approaches. This knowledge is essential for navigating the legal responsibilities associated with developing, deploying, and using AI technologies responsibly.
    \item Ethical Awareness and Reasoning (EAR): This dimension is complex and it is "necessary to comprise many different and complex aspects under one term" \citep{KNOTH2024100177}.  It encompasses the ability to identify, evaluate, and address ethical implications of AI technologies across multiple levels \citep{9844014}. E.g., at the individual level, it includes considerations of safety, privacy and data protection, freedom and autonomy, and human dignity. At the societal level, it involves understanding fairness and justice, responsibility and accountability, transparency, surveillance implications, controllability of AI, impacts on democracy and civil rights, job replacement concerns, and effects on human relationships. At the environmental level, it addresses ethical considerations related to natural resource consumption, energy usage, environmental pollution, and sustainability \citep{9844014}. 
    \item Social Impact Understanding (SIU): This dimension encompasses the ability to comprehend and critically evaluate the multifaceted effects of AI technologies on societal structures and human interactions. Based on research by \citet{KHOGALI2023102232, Ng2023, Almatrafi2024, Miao24}, it includes recognizing both beneficial and detrimental impacts of AI implementation across economic, workplace, and broader social contexts. SIU requires a grasp of the interplay between technological advancements and social, political and economic structures, with particular attention to long-term societal consequences of AI systems on diverse populations and spheres of life.
\end{itemize}

\subsubsection{Delineation from Related Literacies}
Rather than exist in isolation, different literacies coexist and are interconnected with both hierarchical and overlapping elements. For example, the question of whether AIL builds upon Data Literacy or exists as a parallel competency requires careful consideration, while both are considered part of the overarching concept of Digital Literacy. As \citet{Schller2022} argues in their conceptualization of Data Literacy, the ability to understand, interpret, and work with data forms a foundation on which AIL can develop. However, AIL extends these skills to encompass unique dimensions such as human-AI interaction and ethical reasoning about autonomous systems. The relationship between these literacies raises important questions about the prerequisites for digital competence. While \citet[p.113]{Maznev2024} question whether digital competence is a prerequisite for AI competence, \citet{Moravec2024} finds that with growing Digital Literacy, the use of AI tools diversifies, and \citet{Long2020} assert that basic computer understanding is necessary for meaningful AI application, though advanced Computational Literacy is not necessarily required. 

AIL goes beyond Computational Literacy by emphasizing human interaction with AI systems and ethical considerations. Unlike Data Literacy, which focuses primarily on data management and statistical reasoning, AIL involves understanding how AI models process and generate outputs. Furthermore, AIL expands Media Literacy by addressing how AI-driven algorithms shape information flow, recommendation systems, and digital media consumption \citep{Long2020, KitNg2022}. This conceptualization of overlapping literacies is supported by recent major frameworks, such as the OECD's AILit framework, which explicitly maps the interdisciplinary connections between AI Literacy, Media Literacy, Data Science, and Ethics \citep{OECD2025ailit}. Table \ref{tab:literacy-comparison} clarifies the current conceptualizations of the different literacies from an educator's point of view. 

\begin{table}[H]
\centering
\small
\begin{tabular}{p{2cm}p{2.5cm}p{2.5cm}p{2.5cm}p{2.5cm}}
\toprule
\textbf{Aspect} & \textbf{AIL} & \textbf{Media Literacy} & \textbf{Data Literacy} & \textbf{Computational Literacy} \\
\midrule
Definition &  Critically engaging with AI technologies, understanding their societal/ethical/legal implications, and applying them responsibly (see 4.2) & Access, analyze, evaluate, create, and act using all forms of communication \citep{NAMLE2025} & Handling, interpreting, and making decisions based on data \citep{Schller2022} & Understanding programming logic and algorithmic thinking \citep{Maznev2024} \\
\addlinespace
Focus & AI technologies: Interaction and Understanding & Content and communications in society & Raw data and interpretation & Programming and development \\
\addlinespace
Orientation & System-oriented: How AI functions and makes decisions & Content-oriented: Meaning and impact of media & Information-oriented: Value and significance of data & Development-oriented: Creating computational solutions \\
\addlinespace
Features & Critical evaluation, responsible use, awareness of ethical, societal, and legal impact & Contains cultural and creative dimensions (genres, symbols, storytelling) & Focuses on statistical methods also independent of AI & Requires deep understanding of code structures \\
\bottomrule
\end{tabular}
\caption{Comparative Analysis: AIL vs. Related Literacy Concepts}
\label{tab:literacy-comparison}
\end{table}

\begin{figure}[H]
    \centering
    \includegraphics[width=0.7\textwidth]{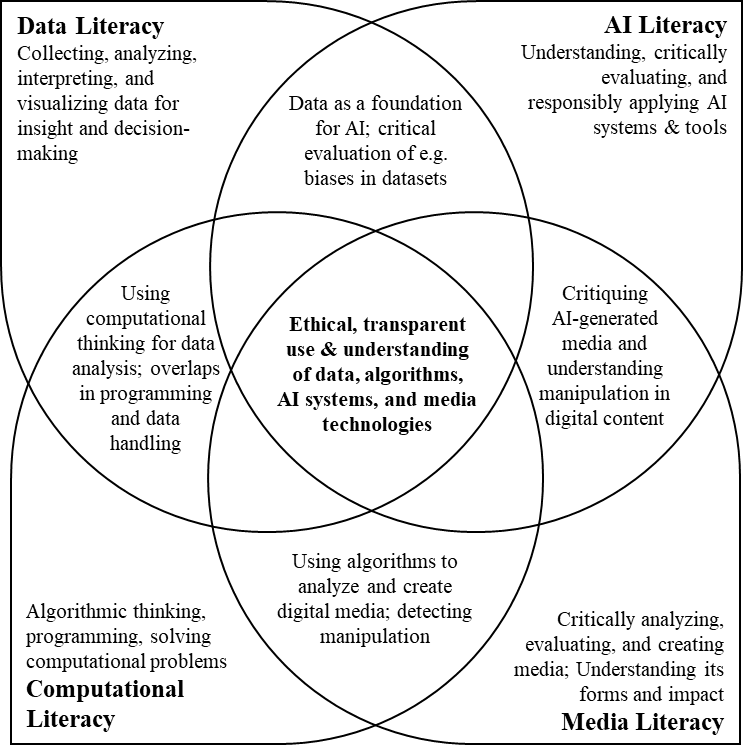}
    \caption{AIL and its overlaps with Media, Computational and Data Literacy}
    \label{fig:delineation}
\end{figure}

Figure \ref{fig:delineation} attempts to illustrate the interrelationships between AIL, Data Literacy, Computational Literacy, and Media Literacy. The diagram's central convergence, emphasizing "ethical, transparent use and understanding of data, algorithms, AI systems, and media technologies," reflects the integrated nature of these literacies in addressing contemporary digital challenges. This aligns with \citet{Klar_Schleiss_2024} observation that attempts to define AI competencies occur through discourse on Data and AI Literacy and are promoted through frameworks from organizations such as UNESCO and the EU. Media Literacy's intersection with AIL, focusing on "critiquing AI-generated media and understanding manipulation in digital content," extends "the ability to access, analyze, evaluate, create, and act using all forms of communication" \citep{NAMLE2025}. Beyond the intersections explicitly delineated between adjacent domains in the diagram, further substantial conceptual overlaps exist. Notably, AI Literacy and Computational Literacy intersect where algorithmic principles underpin the design, analysis, or implementation of AI systems. Likewise, Data Literacy and Media Literacy converge significantly in the critical evaluation of data-driven narratives and visualizations presented within media, as well as in the contextualized communication of data insights through various media channels.

\subsection{Proposing a Synthesized Working Definition of AIL (RQ2)}

Based on the seven core dimensions identified in our literature analysis, we now propose a working definition designed to provide a comprehensive conceptual foundation for the AI Literacy Heptagon. Our review indicates that foundational definitions, such as that by \citet{Long2020}, and influential frameworks from organizations like EDUCAUSE \citep{educause2024ailiteracy} and UNESCO \citep{Miao24}, rightly establish critical evaluation and responsible use as core tenets of AIL. 

However, as our analysis has shown, they often lack an explicit enumeration of all key competency areas, particularly the emerging dimensions of practical integration skills and legal and regulatory knowledge.
Our proposed definition aims to bridge this gap by synthesizing these established concepts with the full spectrum of dimensions identified in our research. As AI technologies and their applications continue to evolve, this working definition is expected to undergo further refinement.

\newtcolorbox{workingdef}{
  colback=gray!10,  
  colframe=gray!50, 
  title=Working Definition,
  fonttitle=\bfseries
}

\begin{workingdef}
AI Literacy in Higher Education refers to the ability to critically engage with AI technologies, understand their societal and ethical implications, and apply them responsibly. It integrates knowledge, skills, and attitudes across technical, applicational, critical thinking, ethical, social, integrational, and legal dimensions, with variations in emphasis depending on the disciplinary context.
\end{workingdef}

This definition positions AIL as a multifaceted construct. Its first sentence aligns with the consensus in the literature, emphasizing critical, ethical, and responsible interaction with AI. The unique contribution lies in the second sentence, which operationalizes this broad goal by explicitly listing the seven core dimensions that form the basis of our Heptagon model. By formally including 'integrational' and 'legal' dimensions, it addresses contemporary challenges that are underrepresented in many existing conceptualizations. Furthermore, the final clause—"with variations in emphasis depending on the disciplinary context"—directly connects the definition to the flexible, adaptable nature of the Heptagon framework, acknowledging that AIL is not a monolithic competency but must be tailored to specific fields of study.

\subsection{Transferring scientific insights on AIL into practice (RQ3): The AI Literacy Heptagon}

\begin{figure}[htbp]
    \centering
    \includegraphics[width=0.9\textwidth]{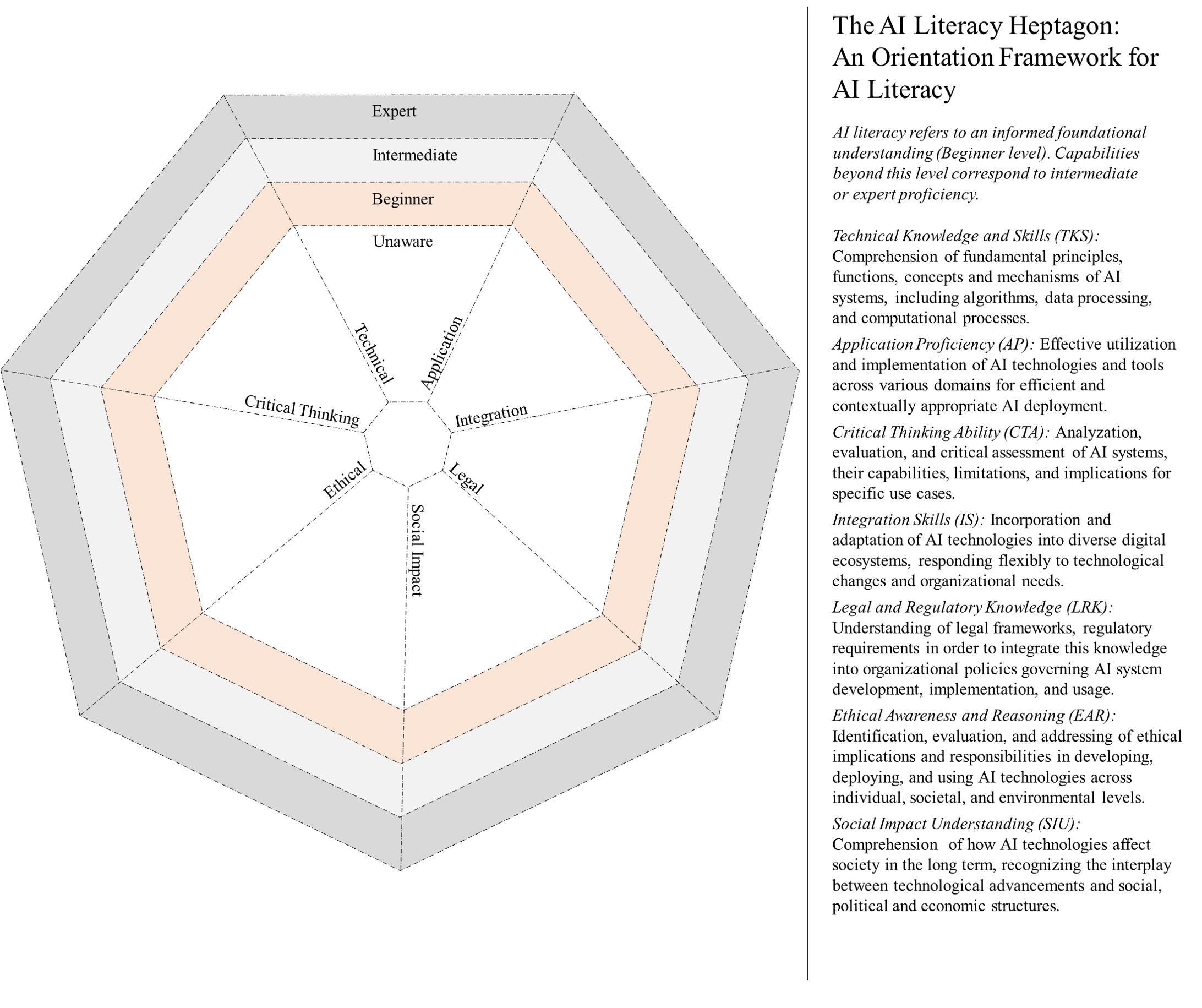}
    \caption{The AI Literacy Heptagon: An Orientation Framework for AIL}
    \label{fig:heptagon}
\end{figure}

To translate this working definition into a practical tool for curriculum development and analysis, we developed the AI Literacy Heptagon (figure \ref{fig:heptagon}). It synthesizes the previous findings into a comprehensive framework that structures the development of AIL in HE. The heptagonal shape represents the seven core dimensions of AIL identified in our literature review, with the aim of reducing the cognitive load while not oversimplifying the construct \citep{SWELLER201137}. The visualization serves multiple purposes:

\begin{itemize}
    \item It illustrates the interconnected nature of the seven dimensions, purposely represented as equal to foster emerging perspectives beyond the dominant technical focus. In particular, legal and regulatory knowledge, often subsumed in ethics, deserves separate consideration, as highlighted by \citep{annapureddy10.1145/3685680}, responding to evolving regulatory landscapes like the EU AI Act.
    
    \item The framework incorporates progression levels derived from Bloom's taxonomy, allowing a structured approach to competency development from basic awareness to Expert-level capabilities, based on \citet{Ng2021}.
    
    \item We distinguish between generic AIL (Beginner level in all dimensions) as the baseline required for all HE students \citep{Laupichler2023, Long2020, Ng2021}, and domain-specific extensions (Intermediate and Expert levels) that should be tailored to particular fields of study \citep{Schleiss2023, cerny13030129, Long2020, KNOTH2024100177}.
    
    \item The adaptability allows disciplines to emphasize different dimensions (e.g., Technical Knowledge in STEM fields; Ethical Reasoning in humanities) while maintaining the multidimensional integrity essential to comprehensive AIL.
\end{itemize}

The four competency levels (Unaware, Beginner, Intermediate, Expert) provide educational stakeholders with clear developmental pathways. Each level corresponds to specific cognitive processes:\\

\begin{tabular}{p{4cm}p{8cm}}
    \textbf{Unaware:} & No awareness of AI concepts, implications, or  applications; lacks knowledge and engagement with AI-related topics \\
    \textbf{Beginner level:} & Basic AI knowledge and competencies for individuals without specialized training, broad, foundational understanding, aligns with remembering and understanding\\
    \textbf{Intermediate level:} & Advanced AI competencies for specialized professionals, encompasses applying and analyzing \\
    \textbf{Expert level:} & Expertise and the ability to innovate and develop new AI solutions, involves evaluating and creating \\
\end{tabular}

\begin{table}[H]
    \centering
    \renewcommand{\arraystretch}{1.2}
    \setlength{\tabcolsep}{5pt}
    \scriptsize
    \begin{tabular}{p{2cm} p{2cm} p{3cm} p{3cm} p{3cm}}
        \toprule
        \textbf{Dimension} & \textbf{Unaware} & \textbf{Beginner (Remembering, Understanding)} & \textbf{Intermediate (Applying, Analyzing)} & \textbf{Expert (Evaluating, Creating)} \\
        \midrule
        \textbf{Technical Knowledge and Skills (TKS)} & No awareness of AI systems. & Identifies and explains basic principles, functions, and mechanisms of AI systems. & Applies and analyzes AI algorithms and data processing techniques in practical scenarios. & Evaluates, develops, and optimizes advanced AI solutions and computational processes. \\
        \midrule
        \textbf{Application Proficiency (AP)} & Does not utilize AI tools. & Recognizes and describes basic AI applications. & Implements and adapts AI technologies to solve practical problems effectively. & Critically evaluates and strategically designs innovative AI applications. \\
        \midrule
        \textbf{Critical Thinking Ability (CTA)} & Not aware of AI implications or limitations. & Recognizes and explains basic capabilities and limitations of AI systems. & Analyzes and critically assesses AI suitability for specific contexts. & Thoroughly evaluates AI systems and proposes advanced solutions to emerging issues. \\
        \midrule
        \textbf{Integration Skills (IS)} & No ability to incorporate AI technologies into personal or professional digital environments. & Recognizes and describes basic opportunities for integrating AI solutions across various contexts (professional, educational, personal). & Implements and analyzes integration strategies, flexibly adapts AI solutions to different application contexts, and responds to technological changes. & Evaluates complex integration scenarios, designs innovative AI integrations, and develops adaptive strategies for incorporating new AI developments across different life domains. \\
        \midrule
        \textbf{Legal and Regulatory Knowledge (LRK)} & No awareness of legal frameworks concerning AI. & Recalls and understands fundamental legal and regulatory frameworks governing AI. & Applies and analyzes regulatory implications in specific AI scenarios. & Evaluates legal risks comprehensively and designs guidelines compliant with evolving AI regulations. \\
        \midrule
        \textbf{Ethical Awareness and Reasoning (EAR) } & Not aware of ethical implications related to AI. & Recognizes and explains ethical principles relevant to AI. & Applies ethical principles and analyzes ethical implications in specific AI scenarios. & Evaluates complex ethical challenges and formulates standards for ethical AI development and use. \\
        \midrule
        \textbf{Social Impact Understanding (SIU)} & Not aware of societal impacts of AI technologies. & Recognizes and explains basic social impacts of AI on human interaction and work environments. & Assesses and analyzes how AI technologies influence society and social structures. & Evaluates complex societal impacts, anticipating future consequences, and develops strategies for responsible AI integration. \\
        \bottomrule
    \end{tabular}
    \caption{Graduated AIL Levels across Key Domains Based on Bloom's Taxonomy}
    \label{tab:ai_competence}
\end{table}

\begin{figure}[H]
    \centering
    \includegraphics[width=1.2\textwidth]{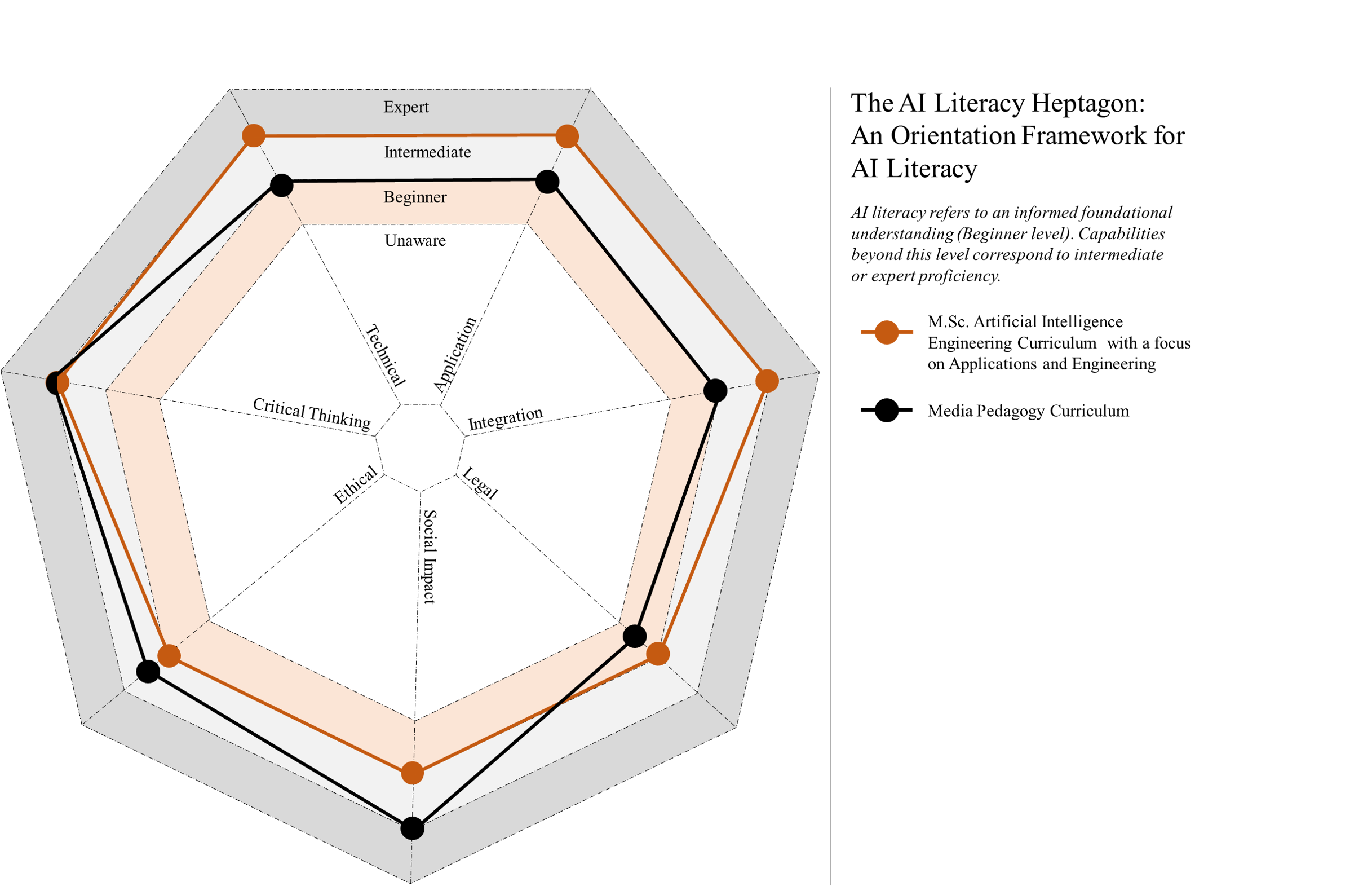}
    \caption{The AI Literacy Heptagon: Example Profiles}
    \label{fig:heptagon_examples}
\end{figure}

Based on two example study plans for the Bachelor's program in AI Engineering (preliminary module catalog), and the Media Pedagogy Curriculum (Teacher Education for Primary and Secondary Schools, Intermediate Schools, and High Schools) \citep{modulkatalog_medienpaedagogik_2024}, the following profiles were developed through an expert-led curriculum mapping process and are depicted in figure \ref{fig:heptagon_examples} and described in more detail in tables \ref{tab:CurriculumMapping1} and \ref{tab:AILiteracyMediaPedagogy}. 

The study programs were chosen to demonstrate that AIL as a compound construct is not an entirely new idea; instead, it builds upon existing grounds. As an example, as media pedagogy involves media literacy, the profile is relatively strong compared to other teacher education programs due to the overlaps between media and AIL, cf. figure \ref{fig:delineation}. Both programs equip students with generic AIL, while each one of them offers different foci depending on their respective domains. Nonetheless, it shows areas to improve, such as that, for example, the Media Pedagogy Curriculum could benefit from more legal and regulatory knowledge in their domain-specific context. 

\section{Discussion}

The results presented in the previous section provide the empirical and conceptual foundation for our proposed framework. The following discussion interprets and critically examines the strengths, weaknesses, and broader implications of the AI Literacy Heptagon. The findings are then connected back to the initial research gaps and explore the consequences of our framework for curriculum development in HE.

Our analysis reveals several strengths and weaknesses of the proposed framework. First, it offers a comprehensive yet manageable conceptualization that balances the need for holistic coverage without overwhelming complexity. Second, the framework incorporates dimensions that are underrepresented in current research, particularly Legal and Regulatory Knowledge (LRK) and Integration Skills (IS), addressing gaps identified in our literature review. 

Beyond guiding curriculum development, the AI Literacy Heptagon offers implications for the design of educational resources and interventions. For instance, it can inform the creation of targeted learning modules addressing specific dimensions or varying proficiency levels across the Heptagon's dimensions, as well as inspire the design of assessment tasks that require students to integrate and apply competencies from multiple AIL dimensions (e.g., ethically evaluating and technically integrating an AI tool for a specific academic task). 

A limitation identified through expert validation is that the visualization lacks sufficient granularity at the expert level. Enhancing detail would provide better guidance for advanced AIL development, but would significantly increase visualization complexity. Furthermore, while our framework conceptually acknowledges that competencies integrate knowledge, skills, and attitudes, this three-dimensional nature is not explicitly represented in the visual model.

The curriculum analysis of Media Pedagogy and AI Engineering programs demonstrated the discipline-specific nature of AIL implementation. As expected, technical programs emphasize Technical Knowledge and Skills (TKS) and Application Proficiency (AP), while humanities-oriented programs focus more on Ethical Awareness and Reasoning (EAR) and Social Impact Understanding (SIU). This finding supports our argument for a flexible framework that maintains core competencies while allowing for disciplinary specialization.

A key contribution of the framework is the explicit inclusion and detailed conceptualization of Legal and Regulatory Knowledge (LRK). The analysis reveals that this dimension is characterized by an interplay between new, AI-specific laws and established, technology-neutral frameworks. This highlights a challenge for HE: AIL curricula must prepare students for a moving target, where legal requirements are evolving. The Heptagon model addresses this by treating LRK as a core competency, encouraging educational programs to move beyond a purely technical or ethical focus and equip students with the necessary knowledge about compliance challenges.

A persistent challenge observed in programs is the disconnect between theoretical knowledge and practical application. While students may develop conceptual understanding of AI principles, they often lack opportunities to apply this knowledge in authentic contexts. This theory-practice gap reflects broader implementation challenges in HE, where AIL initiatives frequently exist as isolated interventions rather than systematic, curriculum-wide implementations \citep{Lambrecht2025}.

The progression model based on Bloom's taxonomy (Unaware → Beginner → Intermediate → Expert) provides educational stakeholders with a structured developmental pathway that makes implementation more manageable. We propose that the Beginner level across all seven dimensions represents the essential baseline of AIL required for all HE students, regardless of disciplinary focus \citep{Laupichler2023, Long2020, Ng2021}.

A limitation of our study is its focus on HE without examining workplace applications; therefore, the transferability of findings to professional contexts requires further investigation. 
While the expert-led curriculum mapping, as detailed in Section 3.2, provides initial insights into the framework's utility, we acknowledge its limitations regarding subjective bias and the absence of direct student competency measurement. This approach represents a foundational step. To achieve a more robust and consensus-driven validation in the future, employing a Delphi methodology is a suggested key direction for further research. Although conducting such an iterative study was beyond the scope of this paper, it offers a path toward enhancing the scientific rigor of the framework.

The Heptagon might support the field of AIL assessment research. Existing instruments such as SNAIL \citep{Laupichler2023} emphasize technical, critical appraisal and application dimensions; more recent questionnaires such as AILQ \citep{https://doi.org/10.1111/bjet.13411} or AICOS \citep{markus2025objectivemeasurementailiteracy} provide broader perspectives. Similarly, psychometric frameworks and research on predictors of AIL are needed to answer questions such as the role of cognitive abilities or educational background for developing AIL \citep{li2025gfactorafactorestablishingpsychometric}.

\section{Conclusion}

This integrative literature review has synthesized current research on AIL in HE, resulting in the AI Literacy Heptagon framework. This structured approach integrates seven core dimensions, providing educators with an adaptable model for curriculum development.

The paper makes several key contributions to the field. First, it provides a systematic delineation of AIL from related concepts such as Media, Data, and Computational Literacy, identifying both overlaps and distinctive features. Second, it implements a progressive competency model based on Bloom's taxonomy that distinguishes between generic AIL (relevant for all students) and domain-specific extensions (discipline-dependent). Third, it identifies and integrates underrepresented dimensions, particularly integration skills and legal/regulatory knowledge, that are increasingly critical in light of evolving regulatory frameworks.

Future research should focus on empirical validation across diverse disciplines, particularly in the humanities and social sciences. Comparative analyses examining AIL manifestations in technical versus non-technical domains would further refine the framework's adaptability. AIL assessment remains a significant challenge for future research and effective transfer to practice. Further necessary work involves undertaking additional qualitative and quantitative research to deepen the understanding and operationalization of specific dimensions, particularly the less-established Integration Skills and Legal/Regulatory Knowledge, and to better characterize the competencies associated with the Expert proficiency level.

Given the rapid development of AI technologies, universities should adopt adaptive curricula that integrate AIL across multiple disciplines. This could be achieved by embedding AIL modules into existing Digital Literacy courses and offering interdisciplinary AI ethics and legal seminars. The AI Literacy Heptagon offers HE institutions a framework for systematically integrating AI competencies into curricula, enabling students to develop the multidimensional knowledge necessary to critically engage with, responsibly apply, and ethically govern AI technologies in their respective fields.

\section{Appendix}

\begin{table}[H]
    \centering
    \renewcommand{\arraystretch}{1.2}
    \setlength{\tabcolsep}{5pt}
    \scriptsize
    \begin{tabular}{p{2cm} p{5.5cm} p{5.5cm}}
        \toprule
        \textbf{Category} & \textbf{Inclusion Criteria} & \textbf{Exclusion Criteria} \\
        \midrule
        \textbf{Thematic Focus} 
        & Explicit focus on AIL or AI competence 
        & Exclusively technical aspects of AI without an educational focus \\
        & Integration of AI in HE 
        & Studies without relevance to HE \\
        & AI-related competencies for students and educators 
        &  \\
        & Interdisciplinary perspectives on AIL 
        &  \\
        \midrule
        \textbf{Context} 
        & HE context or direct applicability to HE 
        & Focus solely on K-12 or vocational education without transferability to HE \\
        & Consideration of perspectives from educators and students 
        & Studies with a highly specific focus lacking generalizable insights on AIL \\
        \midrule
        \textbf{Recency and Relevance} 
        & Publications from 2021–2024 
        & Publications before 2021 (except foundational works) \\
        & Foundational, frequently cited works on AIL (including older publications) 
        & Outdated concepts without current relevance \\
        \midrule
        \textbf{Research Design and Quality} 
        & Empirical studies (qualitative, quantitative, mixed-methods) 
        & Non-peer-reviewed articles or conference proceedings(except foundational works) \\
        & Systematic literature reviews and meta-analyses 
        & Popular science publications without a solid scientific foundation \\
        & Conceptual papers on the definition and operationalization of AIL 
        &  \\
        \midrule
        \textbf{Language} 
        & English and German-language publications
        & Publications in other languages \\
        \bottomrule
    \end{tabular}
    \caption{Selection Criteria for Literature Review}
    \label{tab:selection_criteria}
\end{table}

\begin{table}[H]
\centering
\small
\begin{tabular}{p{5cm}|c|c|c|c|c|c|c}
\hline
\textbf{Authors \& Year} & \textbf{TKS} & \textbf{AP} & \textbf{CTA} & \textbf{EAR} & \textbf{SIU} & \textbf{IS} & \textbf{LRK} \\
\hline
\citet{TokerGokce2024} & x & x & x & x &  &  &  \\
\hline
\citet{Mansoor2024} & x & x & x & x &  & x &  \\
\hline
\citet{Tzirides2024} & x & x & x & x &  & x &  \\
\hline
\citet{folmeg} & x & x & x & x &  & x &  \\
\hline
\citet{Ndungu2024} & x & x & x & x &  & x &  \\
\hline
\citet{cerny13030129} & x & x &  &  &  & x &  \\
\hline
\citet{Schller2022} & x &  & x & x & x & x &  \\
\hline
\citet{Delcker2024} & x & x & x & x & x & x & x \\
\hline
\citet{Tenberga2024} & x & x & x & x &  & x &  \\
\hline
\citet{Sperling2024} &  & x & x &  &  & x &  \\
\hline
\citet{KitNg2022} & x & x & x & x &  &  &  \\
\hline
\citet{Carolus2023} & x & x & x & x & x & x &  \\
\hline
\citet{EU} [AI Act]  &  & x & x & x & x &  & x \\
\hline
\citet{KNOTH2024100177} & x & x & x &  & x & x &  \\
\hline
\citet{Long2020} &  & x & x &  &  & x &  \\
\hline
\citet{Miao24} & x & x &  & x & x &  &  \\
\hline
\citet{Pinski2024} & x & x & x &  &  &  &  \\
\hline
\citet{Al-Abdullatif2024} & x & x & x &  &  & x &  \\
\hline
\citet{Celik2023} & x & x & x & x & x &  &  \\
\hline
\citet{Lintner2024} & x & x & x & x &  &  &  \\
\hline
\citet{Walter2024} & x & x &  & x & x & x &  \\
\hline
\citet{Wang2024} & x & x & x & x &  &  &  \\
\hline
\citet{Knoth2024} & x & x & x &  &  & x &  \\
\hline
\citet{Ng2023} & x & x & x & x & x &  &  \\
\hline
\citet{Hornberger2023} & x & x & x & x & x & x &  \\
\hline
\citet{Almatrafi2024} & x & x & x & x & x & x &  \\
\hline
\citet{Ng2021} & x & x &  & x &  & x &  \\
\hline
\end{tabular}
\caption{Literature Review on AIL Components}
\label{tab:ai-literacy-review}
\end{table}

\begin{table}[H]
\centering
\renewcommand{\arraystretch}{1.2}
\begin{tabular}{p{1cm} p{2.5cm} p{10cm}}
\hline
\textbf{Dim.} & \textbf{Level} & \textbf{Description} \\ \hline

\textbf{TKS} 
& \textbf{Intermediate} 
& 
{coverage of technical aspects through courses like ``Machine Learning,'' ``Deep Learning,'' ``Probabilistic Machine Learning,'' and ``Multiagent Systems.''; progress from foundational understanding to practical application in projects and Bachelor Thesis; combination of theoretical input (lectures) and practical exercises}
\\

\textbf{AP} 
& \textbf{Intermediate} 
& 
{mediation of application skills through project-based courses; Applied Track (``Data Base and Information Systems I'', ``Software Engineering'', and ``Data and Knowledge Engineering'') }
\\

\textbf{CTA} 
& \textbf{Intermediate} 
& 
{includes courses to develop analytical skills (such as ``Linear Algebra'' and ``Analysis I'')}
\\

\textbf{IS} 
& \textbf{Intermediate} 
& 
{Applied Track as well as ``Project AI'' prepare students to integrate AI into environments or contexts}
\\

\textbf{LRK} 
& \textbf{Beginner} 
& 
{``Legal Issues of AI'' provides basic knowledge of legal frameworks governing AI systems}
\\

\textbf{EAR} 
& \textbf{Beginner} 
& 
{``Ethical Aspects of AI'' offers basic coverage of ethical implications}
\\

\textbf{SIU} 
& \textbf{Beginner} 
& 
{e.g. optional module ``Entrepreneurship'' suggests consideration of economic and business impacts}
\\

\hline
\end{tabular}
\caption{Example Curriculum Mapping: Bachelor's program in AI Engineering (Applied Track)}
\label{tab:CurriculumMapping1}
\end{table}

\begin{table}[H]
\centering
\renewcommand{\arraystretch}{1.2}
\begin{tabular}{p{1cm} p{2.5cm} p{10cm}}
\hline
\textbf{Dim.} & \textbf{Level} & \textbf{Description} \\ \hline

\textbf{TKS} 
& \textbf{Beginner} 
& 
{Introduction to informatics and fundamental software development concepts; \textit{Ideas of Informatics} provides students with basic knowledge in computational modeling, development, and analysis of computer programs}
\\

\textbf{AP} 
& \textbf{Beginner} 
& 
{Practical application in media design and digital media didactics; \textit{Design of Digital Learning Media} covers instructional design, creation of digital learning materials, and technical aspects of media production}
\\

\textbf{CTA} 
& \textbf{Intermediate} 
& 
{Critical reflection on media influences and evaluation of media-based teaching strategies; \textit{Media Pedagogical Questions of School Development} explores implications of digital transformation for teachers, students, and educational institutions}
\\

\textbf{EAR} 
& \textbf{Beginner to Intermediate} 
& 
{Ethical considerations in media education and digital learning environments; \textit{Selected Questions of Media Education} focuses on topics such as youth protection, responsible media usage, and ethical dilemmas in digital learning}
\\

\textbf{SIU} 
& \textbf{Intermediate} 
& 
{Analysis of societal and pedagogical impacts of media and digitization; \textit{Selected Questions of Media Didactics} examines the role of digital media in education, Media Literacy, and the changing role of teachers in the digital era}
\\

\textbf{IS} 
& \textbf{Beginner} 
& 
{a practical module allows students to design and implement media-based educational projects, integrating theoretical and practical aspects of digital education}
\\

\textbf{LRK} 
& \textbf{Unaware to Beginner} 
& 
{Basic knowledge of copyright, data protection, and regulatory frameworks; \textit{Design of Digital Learning Media} introduces students to basic legal considerations when producing and using digital media in educational contexts}
\\

\hline
\end{tabular}
\caption{Example Curriculum Mapping: Media Pedagogy Curriculum}
\label{tab:AILiteracyMediaPedagogy}
\end{table}

\bibliographystyle{elsarticle-harv}
\bibliography{1_literatur}

\end{document}